# Density functional theory study on the dihydrogen bond cooperativity in the growth behavior of dimethyl sulfoxide clusters

Natarajan Sathiyamoorthy Venkataramanan,[a,b]* Ambigapathy Suvitha[a] and Yoshiyuki Kawazoe[a]

[a]New Industry Creation Hatchery Center, Tohoku University
6-6-4 Aramaki-aza-Aoba,
Aoba-ku,
Sendai 980 -8579
Japan

[b] Present address
Center for Computational Chemistry and Materials Science (CCCMS),
SASTRA University
Thanjavur – 613  401.  India
Email: nsvenkataramanan@gmail.com; venkataramanan@scbt.sastra.edu
Tel: +91-4362-272946 ; Fax ; +91-4362-264-120

**Graphical Abstract**

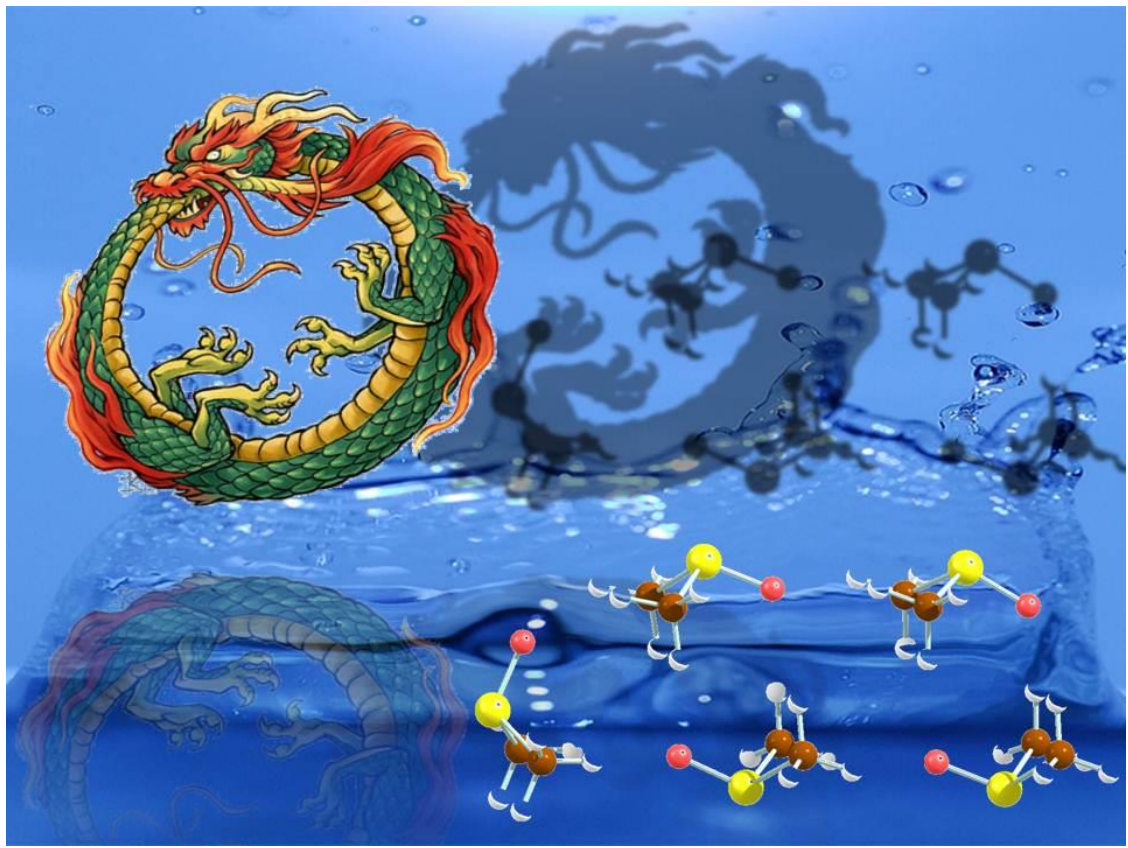

**Highlights**

➢ The ground state geometry of DMSO clusters prefer an ouroboros shape.

➢ σ-hole on the odd numbered clusters helps for the highly directional growth.

➢ AIM analysis shows the existence of intermolecular dihydrogen bonds.

➢ Cooperative effects help to increase the stability of the DMSO cluster.

**Abstract**

We have carried out a density functional theory study on the structures of DMSO clusters and analysed the structure and their stability using molecular electrostatic potential and quantum theory of atoms-in-molecules (QTAIM). The ground state geometry of the DMSO clusters, prefer to exist in ouroboros shape. Pair wise interaction energy calculation show the interaction between methyl groups of adjacent DMSO molecules and a destabilization is is created by the methyl groups which are away from each other. Molecular electrostatic potential analysis shows the existence of σ-hole on the odd numbered clusters, which helps in their highly directional growth. QTAIM analysis show the existence of two intermolecular hydrogen bonds, of type S−O···H−C hydrogen bonds and methyl C−H···H−C dihydrogen bonds. The computed $\rho$ and Laplacian values were all positive for the intermolecular bonds, supporting the existence of noncovalent interactions. The computed ellipticity for the dihydrogen bonds have values > 2, which confirms the delocalization of electron, are mainly due to the hydrogen-hydrogen interactions of methyl groups. A plot of total hydrogen bonding energy vs the observed total local electron density shows linearity with correlation coefficient of near unity, which indicates the cooperative effects of intermolecular dihydrogen H···H bonds.

**KEYWORDS :**

## 1.    Introduction

Dimethyl sulfoxide [DMSO] is an aprotic solvent with high dipole moment (3.96 D) and mixes with water in all propositions at ambient temperature conditions.  Further the presence of the two apolar groups make them soluble in organic media [1]. Due to these physico-chemical properties, DMSO is a very efficient solvent for water-insoluble compounds [2]. Owing to the amphiphilic nature of DMSO and its lower activity on cell membranes they are frequently used as a solvent in biological studies and as a vehicle for drug therapy [3]. Various experimental measurements made for DMSO indicates, that it is a highly associated liquid. The structure of liquid DMSO has been studied earlier using both experimental and theoretical methods [4]. The combined results of neutron diffraction with isotope substitution and molecular dynamics, on the liquid DMSO indicate a small disorder in molecular association [5]. X-ray and neutron diffraction studies had satisfactorily elucidated the intermolecular geometry of DMSO in the liquid state. Furthermore, liquid DMSO is relatively unstructured, with a slight preference for antiparallel ordering of the S–O dipoles [6].

Polarizability and molecular radius measurement made on DMSO indicate that they are not spherical molecules [7]. Figueroa *et al.* using infrared spectra have concluded DMSO molecules in $CCl_4$ are associated as dimers by dipole forces in the concentration range 0.1~0.3 M, while at higher concentrations, associated chainlike structures were suggested [8]. Fini and Mirone present the isotropic and the anisotropic spectrum of liquid DMSO in the S–O stretching vibration region, which are explained by the presence of molecular clusters [9]. The temperature dependent [1]H NMR spectra and light scattering measurement show that DMSO exists as clusters at low temperature [10]. Furthermore, applying electric field result in the formation of larger aggregates [11]. High pressure Raman spectroscopic studies conclude that pressure favours the formation of aggregates and results in a more ordered local structure compared to the structure at ambient pressure [12]. Very recently, Singh *et al* reported the concentration dependent Raman spectra of neat DMSO and its binary mixtures with water/methanol. The spectral analysis shows that DMSO dimer exists even at low concentration [13].

Dielectric relaxation experiment carried out by Shikata and Sugimoto showed the existence of dimeric DMSO association in the range of 50 MHz – 50 GHz [14]. Moreover, the concentration of dimer was governed by the chemical equilibrium, that depends on the concentration of DMSO and the solvent species. The equilibrium constant for dimer was found to be $20-40$ $M^{-1}$ for DMSO solvent [15]. Gajda and Katrusiak studied the structure of DMSO using single crystal XRD at high pressure. At 140 MPa, the crystal structure exists in α-phase and is dominated by electrostatic attraction and the CH···O hydrogen bonds was responsible for the dimer and chain structure. At a pressure of 540 MPa, the electrostatic forces and half of the dipole-dipole centres between S=O groups are broken and H···H contacts between electropositive atoms are formed in the β – phase [16].

Theoretical interest in the molecule originates from a long-standing controversy of how to correctly describe the sulphur–oxygen bond which is responsible for its high dipole moment [17]. Furthermore, DMSO is the only pyramidal molecule among similar widely used organic compounds. The structure of DMSO molecule has been investigated by many explorations using various theoretical methods [18]. Extensive classical simulations using semi-empirical Coulomb and 6-12 Lennard-Jones interactions potential has been carried out and no ordered molecular association has been found in these molecules [19]. Varnali showed using PM3 calculations, the existence of trimers in linear fashion, which represents segment for higher polymers [20]. Recently, Politzer et al, have carried out DFT studies on the DMSO dimer molecules to understand the role of $\sigma$ – hole bonding in DMSO and its related molecules [21,22]. Recently Venkataramanan studied the nature of interaction between the DMSO and water molecules and observed the existence of cooperativity in intermolecular hydrogen bonding, which has led to the strong complex formation [23,24]. Here, we report on the first and systemic study on the structural growth of $(DMSO)_n$ clusters (n = 1 − 13) using the density functional theory, to get insight into the factors that are responsible for the stability of these clusters.

## 2.      Computational Details

All density functional theory calculations reported here were carried out using the Gaussian 09 suite of programs [25]. The geometries of the DMSO clusters were optimized using various density functionals (DFs) and 6-311++G (d,p) basis set. For search of lowest energy geometry, we used the dispersion corrected B3LYP-D3 functionals [26]. It is well known that Grimme's D3 correction has been parameterized for use with Karlsruhe quadruple-$\zeta$ basis set [27]. Recently, Hostaš and Řezáč observed that the accuracy of interaction energies and geometries could be achieved using the DZVP basis sets while using the dispersion-corrected DFT-D methods [28]. In the present calculation we have used the 6-311++G(d,p), due to the use of large clusters with more than 100 atoms. Thus, the use of 6-311++G(d,p) basis set in these calculations is justified as a compromise between reliable results and reasonable computational cost.   Further, we used the hybrid B3LYP, meta-hybrid GGA M0-52X, M0-62X and dispersion corrected B97D and M06-2X-D3 level of theory and optimized the structures to identify the lowest energy structure in the cases where we observed low lying isomers with 5 kcal mol$^{-1}$ higher in energy [29-32]. In addition to this, we have also carried out geometry optimization at second order Møller-Plesset perturbation (MP2) level for all small size lowest energy clusters (n < 6) and obtained energy from the B3LYP-D3 optimized geometries for medium and large size clusters to compare the accuracy of the present day density functionals. All the structures reported here are fully optimized without any geometrical constraints. Harmonic vibrational frequencies were carried out and the reported structures were found to have no negative frequencies, indicating that they are all in the minima in the potential energy surface and are not the saddle points. The basis set superposition error (BSSE) was calculated at the B3LYPD3/6-311++g(d,p), B97D/6-311++g(d,p) and M0-62X-D3/6-311++g(d,p) level using the full counterpoise procedure using the optimized geometry with the respective functional [33].

Total binding energies for the (DMSO)$n$ clusters are calculated as

$$BE = E(DMSO)n - nE(DMSO) \tag{1}$$

While incremental binding energy for the succesive DMSO addition is computed as

$$IBE = E(DMSO)n - E(DMSO)n\text{-}1 - E(DMSO) \tag{2}$$

where the $E$(DMSO) is the monomer energy and $E$(DMSO)$n$ and $E$(DMSO)$n\text{-}1$ are the energy of the n$^{th}$ and n-1$^{th}$ DMSO clusters, which are determined at the same level of theory. Free energy contributions were obtained from frequency calculations using B3LYP-D3/6-311++G(d,p) level of theory at 250 K, 273 K, 298 K and 313 K.

In order to gain further insights into the bonding that exists in the DMSO clusters, topological analysis has been performed using AIMALL package and the corresponding wave functions were generated at the B3LYP-D3/6-311++G(d,p) level of theory [34]. The Wave function analysis–surface analysis suite (WFA-SAS) program was employed to calculate quantitative electrostatic potential and to visualize the 3D surface [35]. MESP analysis are commonly carried out on van der Waals molecular radius and are analysed at the 0.001 electrons/Bohr$^3$ isodensity surfaces.

## 3.    Results and discussions

### 3.1.    Geometry

The ground state geometry of the DMSO molecule is shown in supporting information Fig. S1, and the geometrical parameters are provided in supporting information Table S1. We have compared the calculated geometrical parameters at various levels of theory. These results indicate that higher level of theory and basis sets are important for accurate determination of bond parameters, while the B97D/6-311++G(d,p) level of theory predicts the dipole moment closer to the experimental value. Since the growth motifs of DMSO clusters are unclear, we have made an extensive search to find the minimum energy structures of dimers, trimers and tetramer clusters in two ways: (i) by considering the possible structures reported in the previous papers for $H_2O$, $NH_3$, $CH_3CN$, and formamide clusters and (ii) by placing the DMSO at various adsorption sites, which leads to numerous stable isomers [36-40]. The ground state search for the hexamer clusters are done by introducing one DMSO molecules at various adsorption sites on the ground and first low-lying isomers of the pentamer cluster and the obtained structure is subjected to optimization. For the initial geometrical optimization of the hexamer clusters, we used the B97D/6-31+G(d) level of theory. The obtained isomers are reoptimized with the dispersion corrected B3LYP-D3 functional. The isomers which have an energy difference of 5 kcal mol$^{-1}$ than its low-lying isomer are reoptimized with various functional to identify their relative position.

The ground state geometry of the DMSO dimer shown in Fig. 1(a). It has antiparallel orientation, in which the oxygen atoms of the DMSO are pointing towards the hydrogen atoms of the methyl group in the other DMSO molecule. The distance between the oxygen atom and the hydrogen atoms was calculated using the B3LYP-D3/6-311++G(d,p) level was found to be 2.240 Å, indicating the existence of intermolecular hydrogen bonding between DMSO molecules. The structure shown in the supporting information Fig. S2(a), is 1.80 kcal mol$^{-1}$ higher in energy, in which oxygen atom bind with hydrogen atoms of the methyl group. In the third highest energy isomer, the DMSO molecules were in parallel orientation and oxygen atom in one DMSO is 2.202 Å away from the methyl hydrogen of other DMSO molecule. The third highest energy isomer shown in Fig. S2(c), which is 3.87 kcal mol$^{-1}$ higher in energy, has a longer S-O bond distance compared to the lowest energy structure. The fourth lowest energy structure shown in Fig. S1(d), has the antiparallel configuration and was 4.29 kcal mol$^{-1}$ higher in energy than the lowest energy isomer. Apart from these structures shown in the supporting information Fig. S1(a)–(d), various other initial geometries were studied, which after the geometry optimization using the B3LYP-D3 functional ended to the stationary points shown in Fig. S1.

The structure of stable, lowest energy trimer clusters is shown in Fig. 1(b), and the other low-energy structures are provided in Fig. S3. The ground state geometry for the trimer was found to possess an ouroboros shape in which each of the DMSO molecules are held by other DMSO molecules. The second isomer, shown in Fig. S2(a), has two DMSO molecules in antiparallel configuration and the other DMSO is held by hydrogen bonding, and is found to be 2.68 kcal mol$^{-1}$ higher in energy. Since the Fig. 1(b) and Fig. S3(a) are very close in energy, we used B3LYP, M05-2X, M06-2X, and M06-2X-D3 functionals to optimize the two geometries. By all functionals the geometry in Fig. 1(b) was found to be more stable than Fig. S3(a). The third highest energy isomer was 4.41 kcal mol$^{-1}$ higher in energy than the lowest energy isomer. The fourth highest energy isomer has a butterfly like structure and was 7.10 kcal mol$^{-1}$ higher in energy than the lowest energy isomer.

The lowest energy and the seven low-lying structures along with important bond parameters for the tetramer are shown in Fig. 1(c) and Fig. S4 respectively. The search for the ground state low-lying isomers for tetramer resulted in an elliptical shaped structure as shown in Fig. 1(c). The S–O···H distances in this structure varies between 2.189 to 2.369 Å. From the bond parameters, it is evident that the all the DMSO molecules are not held with same binding energy in the lowest energy tetramer cluster. The next two low-lying isomers are formed by the interaction of two DMSO antiparallel structures with the other two DMSO molecules in similar configuration, but are rotated and are 0.17 and 0.28 kcal mol$^{-1}$ higher in energy. The third and fourth lowest energy isomers are formed by the interaction between one DMSO molecules with an ouroboros shaped structure formed between three DMSO molecules. It is found that the linear isomer shown in Fig. S4(i), is 1.04 kcal mol$^{-1}$ higher in energy. It is worth to point out here, that linear isomers are reported as the lowest energy structures through classical semi-empirical simulations [20]. The tetramer with stacked configuration shown in Fig. S4(j), was 1.23 kcal mol$^{-1}$ higher in energy. Based on the above discussion, it may be inferred that DMSO molecules prefer to form ouroboros shaped clusters.

In Fig.1(d), we report the lowest energy isomer for the pentamer cluster and in Fig. S5(a–g) we show the first seven low lying energy isomers for pentamer clusters. Apart from the species listed in Fig. S5, several other isomers were located during the structural search using the B97D/6-311++G(d,p) level of theory. However, such structures upon re-optimization using the dispersion corrected hybrid B3LYP-D3 functional gives rise to the structures provided in Fig. S5. The second lowest energy isomer was a fused structure of a dimer and trimer cluster, which is 4.16 kcal mol$^{-1}$ higher in energy than the lowest energy isomer. The isomer with one DMSO interacting with a tetramer cluster was 5.77 kcal mol$^{-1}$ higher in energy, while the linear isomer shown in Fig. S5(i), is 16.33 kcal mol$^{-1}$ eV higher in energy. These results clearly indicate that only the cyclic ouroboros shaped structures are more stable than linear and other random structures.

The possible DMSO isomers for hexamer size are obtained by adding one, two and three DMSO clusters, on the lowest energy pentamer, tetramer and trimer clusters, respectively from various directions. Further, we have considered the stacked conformer, linear and the ouroboros structures. The low-lying isomers are shown in Fig. S6, and the lowest energy isomer is shown in Fig. 1(e). Among them the linear hexamer cluster was found to the least stable, while the ouroboros structure was found to be lowest energy isomer. The heptamer to 13mer cluster consists of 70 to 130 atoms. Therefore, the detailed exploration of the potential energy surface is very computer intensive and expensive. Further, the number of conformers possible for the heptamer and larger size clusters could be much higher. Thus, taking account of the remarkable stability of the dimer to hexamer clusters with ouroboros structure, we have considered only the ouroboros conformations for the higher order clusters. The putative global minima for n = 8–13 obtained at the B3LYP-D3/6-311++G(d,p) level of theory is shown in Fig. 2. In the optimized configurations of the higher order, the DMSO clusters with even number of DMSO molecules were found to have a symmetric configuration, with the central unit in such systems have multiple dimers. On the other hand, structures with odd number of DMSO molecule, have one DMSO molecule which clips the central unit.

The calculated average S-O and SO···H bond lengths for the clusters are shown in Fig. 3. We observed a monotonous increase in the S-O bond length with the cluster size. On the contrary, an odd-even oscillation in the O···H hydrogen bond length was observed with the increase in the cluster size. Further, the computed dipole for the odd numbered clusters has almost 4.20 μ, which is close to the dipole moment of a free monomer molecule. These findings support the existence of an unpaired DMSO monomer in the odd numbered clusters.

### 3.2. Energetics' and Choice of Functionals

The total binding energy, average binding energy per DMSO and the incremental binding energy for the DMSO clusters are provided in Table 1. The total binding energy of the clusters increases linearly with the cluster size, with almost perfect linear correlation and a regression of 0.999 (See Fig. S7). It should be pointed that the average binding energy per DMSO has increased by 100 % from the dimer to 13 mer. The computed average interaction energy per DMSO (Fig. S8) can be fitted with a second-order polynomial equation with a regression of 0.968, and from the value of intercept, the value of interaction energy at infinite length chain has been obtained as 0.241 eV. Previously, such behaviour has been observed on the cooperativity effect of H-bonding chains of 4-pyridone, HCCBeH and aza-borane chains [41-43]. These observation shows the existence of cooperative effect during the bonding of DMSO clusters.

To better understand the cooperativity in ouroboros clusters, we have computed the pairwise energies for the trimer to heptamer clusters. The counterpoise corrected pairwise interaction energies for the trimer to heptamer clusters are provided in Table 2 and the numbering pattern to identify the pair is provided in Fig. 1. In the trimer cluster, we observe a large stabilization between the (1,3) dimeric pair and relatively less stability in (2,3) dimeric pair. In the tetramer cluster, a larger interaction is observed between (1,3) and relatively a weak interaction between the pairs (1,2), (1,4), (2,3) and (3,4). We also observe a destabilizing effect between the pair (2,4) which are separated by the stable pair (1,3). In the stable pentamer, we observed the existence of strong interactions in the pairs (1,3) and (4,6). It should be pointed out here, that these pairs exist in anti-parallel configuration. A relatively weak interaction is observed between the pair (1,2), (1,6), (2,3), (3,4), (4,5) and (5,6), which exists with a near linear configuration. A small destabilization is observed between the pairs (1,4), (1,5), (2,4), (2,5), (2,6), (3,5), and (3,6) which are far away from each other. Such a destabilization factor has been observed in the crystal at 2.4 GPa pressure [16]. In the heptamer, we observed three strong interaction (1,2), (3,7) and (4,6) arising out of dimeric pairs, which exist in anti-parallel configuration and a weak interaction between the near linear configuration. Similar to the hexamer, the destabilization arises due to the pair which are far away from each other. Thus, in these structures there exists cooperativity between dimeric pairs.

The computed incremental binding energy (IBE) for the lowest energy structures shows DMSO dimer has a very high value and trimer to have the least value. The incremental binding energy as a function of cluster size is shown in Fig. 4, from which it is evident that for clusters with n > 3, have an odd-even oscillatory pattern. The odd clusters have higher IBE, which indicates that they are more stable than the even numbered clusters. This could be due to the clipping of the DMSO molecule to the dimeric pattern. Recently Bakó and Mayer have shown that in water molecules the dipole moment enhances with the cluster size [44]. In the present cause, we observed almost odd-even oscillation for larger size DMSO clusters, when the dipole moment is plotted as a function of cluster size (See Fig. S9). Such a pattern is expected as there exists dimeric structures in even numbered clusters, cancelling each other's dipole, while the odd clusters have an extra terminal DMSO molecule whose dipole is envisaged in the calculation.

To understand the effect of functionals on the binding energy we have them at various levels of theory, by optimizing the clusters at the same level of theory. The computed binding energy for the hybrid functional B3LYP, hybrid meta-GGA functionals M05-2X and M06-2X, dispersion corrected functionals B3LYP-D3 and M05-2X-D3 and the MP2 method are provided in Table 1. From the Table 1, we can see that the MP2 values matches well with the dispersion corrected B3LYP-D3 functional. The use of hybrid meta-GGA functionals (M05-2X and M06-2X) did not improve the binding energies even after the basis set superposition correction. While the pure hybrid B3LYP functional, provides a low binding energy. Thus, the use of dispersion correction is essential in the accurate estimation of binding energies in these clusters. Furthermore, the use of BSSE correction increases the BE values. Prior studies on the molecular aggregates have shown that the counterpoise correction provides error as the fragments within the complex is improved by the basic functions, while such extension is not possible in the calculation of the isolated fragments [44]. Recently, Grimme *et al* showed that incorporation of dispersion for geometrical optimization provides better results than performing BSSE correction [45]. Thus, to estimate the binding energy values of DMSO clusters, extreme care has been taken on the usage of functionals.

### 3.3. Nature of Interaction

### 3.3.1. Quantitative molecular electrostatic potential

To understand the nature of interaction between the molecules in the DMSO clusters, we have performed molecular electrostatic potentials (MESP) and quantum theory of atoms-in-molecular (QTAIM) analysis. MESP has been widely used as a tool to understand the effective localization of electron density in a molecule as well as for characterizing the nature of interactions between molecules [46,47]. The MESP mapped surface together with various positive and negative potentials along with their values, designated as $V_{s,max}$ and $V_{s,min}$ respectively for the monomer, dimer, tetramer, pentamer and 13 mer are shown in Fig. 5, and for other minimum energy clusters are shown in supporting information Fig. S10. In the DMSO molecule, the $V_{s,min}$ values was observed on top of the oxygen atom with $V_{s,min}$ value of -48.86 kcal mol$^{-1}$. The positive values were observed on the hydrogen atoms with a $V_{s,max}$ of 20.14 kcal mol$^{-1}$. We also observe a positive potential designated as σ-hole by Politzer and co-workers, closer to the sulphur atom and adjacent to the two methyl hydrogens with a $V_{s,max}$ of 30.62 kcal mol$^{-1}$ [48].

In the DMSO dimer cluster, the $V_{s,min}$ and $V_{s,max}$ values closer to the oxygen atom and on the hydrogen atoms get reduced to -40.81 and 16.71 kcal mol$^{-1}$ compared to the DMSO molecule. In addition, the σ-hole value observed got reduced to 23.49 kcal mol$^{-1}$.

This reduction in positive and negative charge can be attributed to the electrostatic interaction, which provides additional stability to the molecule. Further, the presence of positive charged regions on the methyl hydrogen and the existence of σ-hole, helps for the further highly directional electrostatic interaction. In the DMSO timer molecule, we observe a $V_{s,min}$ with values of -44.13, -41.03 and -22.89 kcal mol$^{-1}$ on the oxygen atoms. Furthermore, we observe only two σ-hole with values 26.54 and 23.86 kcal mol$^{-1}$. This indicates that one of the three DMSO molecule is loosely bound compared to the other two molecules. In the tetramer, we observed two $V_{s,min}$ with values -40.40 kcal mol$^{-1}$ on the oxygen atoms of the terminal DMSO molecules, while two DMSO molecules at the centre have $V_{s,min}$ values of -10.80 kcal mol$^{-1}$. Furthermore, we observed only two σ-hole with values 23.08 kcal mol$^{-1}$ on the terminal DMSO molecule. This can be envisaged, as the existence of electrostatic interaction of the central DMSO dimer with the other DMSO molecules in an intermolecular manner. In the pentamer, we observed two $V_{s,min}$ values of -43.94 and -40.66 kcal mol$^{-1}$ and the σ-hole on the oxygen atoms and between the hydrogen atoms respectively of the clipping DMSO molecule. In the larger DMSO clusters with n > 6, we do observe a similar pattern of existence of $V_{s,min}$ and σ-hole on the clipping DMSO molecules, which helps the highly directional growth of the DMSO clusters.

### 3.3.2. QTAIM analysis

The quantum theory of atoms in molecules (QTAIM) has been extensively applied to classify and understand bonding interactions in terms of quantum mechanical parameter such as electron density (ρ), the Laplacian of the charge density ($\nabla^2\rho$), total energy density (H(r)) (H(r)= local gradient kinetic energy density G(r) + local potential energy density

V(r)) [49,50]. In an effort to shed some light on the nature of interaction in the DMSO clusters, we applied the QTAIM methodology. The molecular graphs for dimer, trimer, tetramer, pentamer and 13 mer clusters are shown in Fig. 6, and for all other clusters the molecular graphs are provided in the Fig. S10. The sum of the calculated charge density ($\rho$), the Laplacian of the charge density ($\nabla^2\rho$), total energy density (H(r)) and the ratio of -G(r)/V(r) at the BCP's for the intermolecular bonding are provided in Table 3, while the individual intermolecular QTAIM parameters are provided in the Table S2–S13.

It is interesting to observe the presence of inter and intramolecular bonding critical points (BCPs), ring critical points (RCPs) and cage critical points (CCPs) in the DMSO clusters, while the DMSO molecule exists with only intramolecular BCPs. This indicates, there exists intermolecular interactions which generates the RCPs and BCPs. It is evident from the molecular graphs, the intermolecular bonds are composed of S-O···H hydrogen bonds and methyl H···H dihydrogen bonds. From the molecular graphs, one can notice that the dihydrogen at the mid of the DMSO clusters are found to have bifurcated bonds. A deeper insight in the molecular graphs, shows that some of the intermolecular bond paths, especially that of intermolecular hydrogen-hydrogen bonds between the DMSO molecules are not straight lines. The curved intermolecular BCPs indicates the existence of strain as previously observed in cyclic organic molecules [51]. With the increase in the number of DMSO molecule, we observe the increase in the number of BCPs, RCPs and CCPs. The number of CCPs were found to be one less than the cluster size, while the RCPs were almost equal to the number of BCPs that exist in the cluster.

The measure of electron density at the BCP provides the information about the nature and strength of bonding between the molecules in any complex [52]. The sign of the Laplacian is determined by energy that dominates in the bonding zone. A positive $\nabla^2\rho$ indicates that kinetic electron energy density G(r) is greater than the potential electron energy density V(r). Such an interaction is observed in closed shell systems or a noncovalent interaction, where the depletion of electronic charge along the bond path is observed. A negative value of $\nabla^2\rho$ at BCP indicates a covalent bond. Besides the above terms, if the ratio of –G(cp)/V(cp) > 1, then the interaction can also be classified as noncovalent.[38] The ellipticity is expressed by $\varepsilon_b = \lambda_1 / \lambda_2 - 1$, where $\lambda_1$, $\lambda_2$, and $\lambda_3$ are the eigenvalues of the Hessian of the electron density. The ellipticity at BCP measures the extent to which the distribution of the electron density is elongated [53].

The computed $\rho$ values are all positive for the S-O⋯H BCPs and the value lies in the range 0.0136–0.010 a.u. The observed $\rho$ values are close to the previously reported hydrogen bonding's [52,53]. The $\rho$ value which reflects the strength of bonding is high for the even number clusters, indicating the high strength of interaction between the dimeric pairs. The $\rho$ values for the dihydrogen H⋯H intermolecular BCPs lie in the range of 0.004 – 0.003 a.u. Interestingly, the $\rho$ values for the dihydrogen H⋯H intermolecular BCPs for the terminal DMSO are higher than for the inner molecule, however, there exists bifurcated dihydrogen bonds in the inner molecules. The Laplacian values for all the studied clusters are all positive, supporting the existence of noncovalent interactions. Furthermore, the ration of -G(r)/V(r) for the clusters lie in the range of 1.141–1.454, and hence can be classified under the noncovalent interactions. On the contrary, the ellipticity values for the intermolecular bonds decreases with the increase in the strength of interaction. For the O⋯H bonds the ellipticity are always less than one, while the for the dihydrogen H⋯H intermolecular BCPs the ellipticity values are > 2. This confirms the delocalization of electron occurs mainly due to the H···H interaction. This conclusion was supported by the fact that the computed pairwise energy in Table 2, shows that there exists an attraction between the nearby methyl hydrogen bonds.

For the systems with pure noncovalent interaction, the correction between the total local electron density at the BCPs and the total binding energies vary linearly. To test the above supposition, we have plotted the total electron density *vs* the total binding energy of the cluster. A linear dependence was observed with correlation coefficient of 0.998. The correction of local electron densities of S-O··· H bond *vs* the total binding energy was also linear with a correction coefficient of 0.998. Furthermore, the plot of local electron densities at the H···H bond vs the total binding energy was also leaner with a correlation coefficient of 0.995. We computed the hydrogen bond energy $E_{HB}$ ($E_{HB}$ = ½ V(r)) and observed that the S-O···H bonds have an average strength of 2.20 kcal mol$^{-1}$, whereas the H···H have an average strength of 0.50 kcal mol$^{-1}$. A plot of total hydrogen bonding energy *vs* the observed total local electron density shows linearity with correlation coefficient of near unity. This indicates the intermolecular dihydrogen H···H bonds shows cooperative effects and contribute significantly to the stability of the DMSO cluster.

### 3.5. Conclusions

In this work, we have carried out a density functional theory study on the structures of DMSO clusters and analysed the structure and their stability using molecular electrostatic potential and atoms-in-molecules analysis. The ground state geometry of DMSO molecules prefer to exist in ouroboros shape. Linear shaped and stacked clusters were found to be less stable compared to the ouroboros structures. The total binding energy of the clusters increases linearly, while the computed average interaction energy per DMSO can be fitted with a second-order polynomial equation with a regression of 0.968. Pair wise interaction energy calculation shows that there is an interaction between methyl groups of adjacent DMSO molecules and a destabilization is observed with methyl groups which are away from each other. The use of dispersion correction along with B3LYP functional was found to be essential in the accurate estimation of binding energies in the DMSO clusters.

Molecular electrostatic potentials show the existence of σ-hole on the odd numbered clusters, especially on the clipping DMSO molecule, which helps the highly directional growth of the DMSO clusters. QTAIM analysis show the existence of two intermolecular hydrogen bonds, of type S-O···HC hydrogen bonds and methyl C-H···H-C dihydrogen bonds. The computed ρ and Laplacian values were all positive for the intermolecular bonds, supporting the existence of noncovalent interactions. The ration of -G(r)/V(r) for the intermolecular bonds line in the range of 1.141 – 1.454, by which can be classified under the noncovalent interactions. The computed ellipticity for the dihydrogen bonds have values > 2, which confirms the delocalization of electron occurs mainly due to the methyl-methyl interactions. A plot of total hydrogen bonding energy vs the observed total local electron density shows linearly with correlation coefficient of near unity, which indicates the cooperative effects of intermolecular dihydrogen H···H bonds towards the stability of the DMSO cluster.

**Acknowledgements**

NSV thanks the DST for a partial support though project (SB/S1/PC-047/2013). We acknowledge the Center for Computational Materials Science at the Institute for Materials Research for use of the Hitachi SR16000 (model M1) supercomputer system.

**Appendix A. Supplementary data**

Low lying geometries of DMSO clusters, atomic numbering scheme, total interaction energy as a function of clusters size, incremental binding energy as a function of cluster size, and molecular electrostatic potential for DMSO clusters are available in the supporting information data. Supplementary data associated with this article can be found in the online version, at doi: http://dx.doi.org/10.1016/j.molliq.2017.xx.xxx.

**Figure Legend:**

**Figure 1.** Optimized geometries of DMSO (a) dimer (b) trimer (c) tetramer (d) pentamer (e) hexamer and (f) heptamer along with their numbering scheme.

**Figure 2.** Optimized geometries of DMSO (a) octamer (b) nanomer (c) decamer (d) 11 mer (e) 12 mer and (f) 13 mer.

**Figure 3**. Plot of bond distance of S−O and S−O···H as a function of cluster size.

**Figure 4**. Plot of computed incremental binding energy as a function of cluster size.

**Figure 5**. Molecular electrostatic surface potentials mapped on corresponding 0.001 au electron density isosurface of DMSO (a) monomer (b) dimer (c) trimer (d) tetramer (e) pentamer and (f) 13 mer along with the observed $V_{s,max}$ and $V_{s,min}$ values.

**Figure 6**. Molecular topography analysis for the DMSO (a) monomer (b) dimer (c) trimer (d) tetramer (e) pentamer and (f) 13 mer.

**Figure 7**. Linear variation of (a) binding energy with total electron density (b) binding energy with electron density at S−O···H BCPs (c) binding energy with electron density at dihydrogen H···H BCPs (d) hydrogen bonding energy $E_{HB}$ with total electron density.

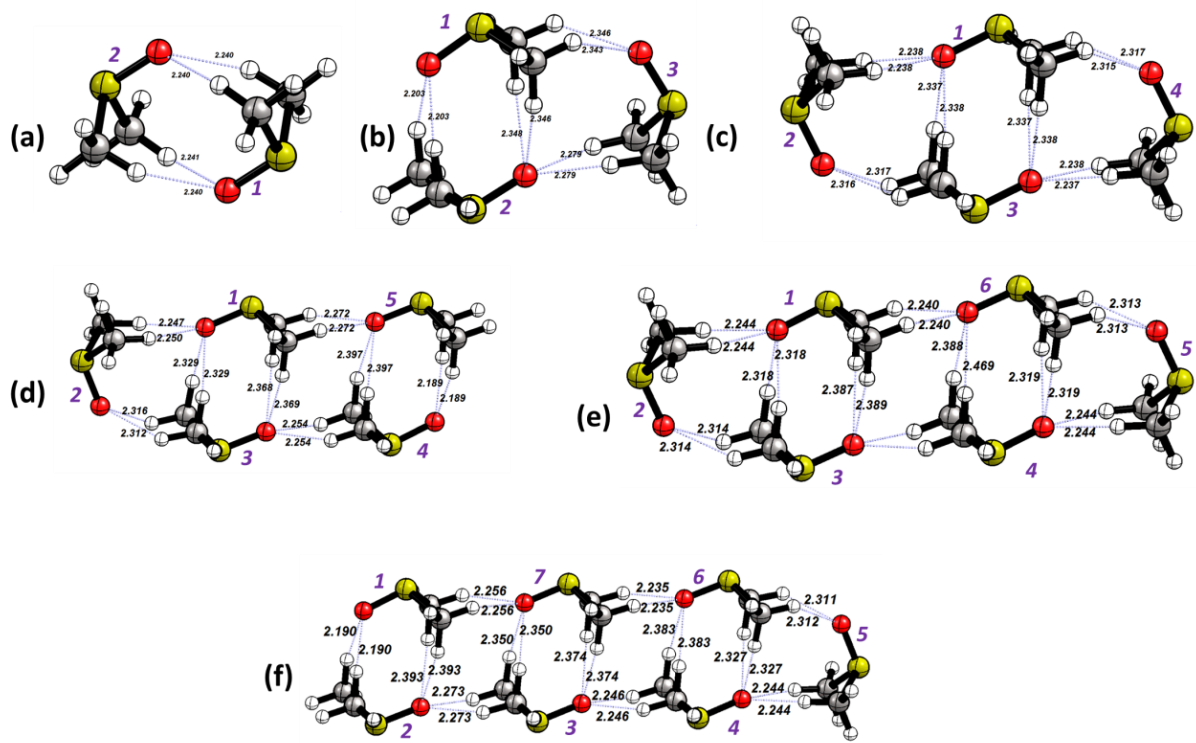

**Figure 1.**

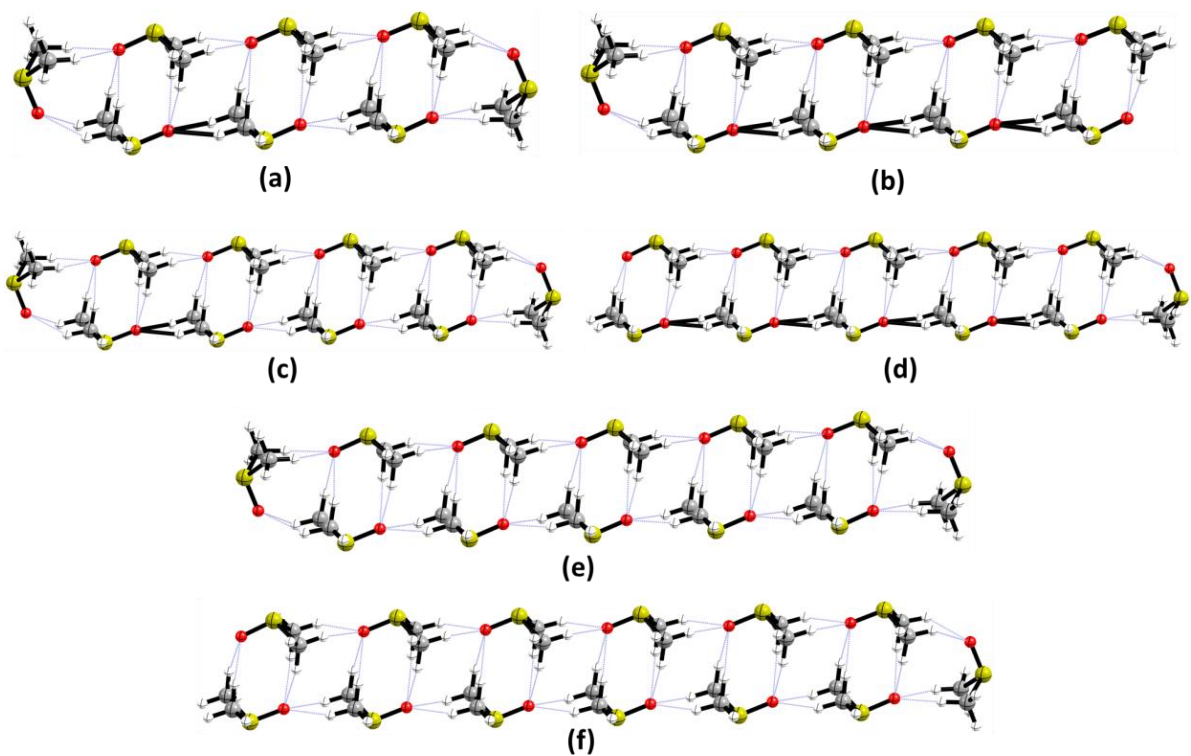

**(a)**

**(b)**

**(c)**

**(d)**

**(e)**

**(f)**

**Figure 2.**

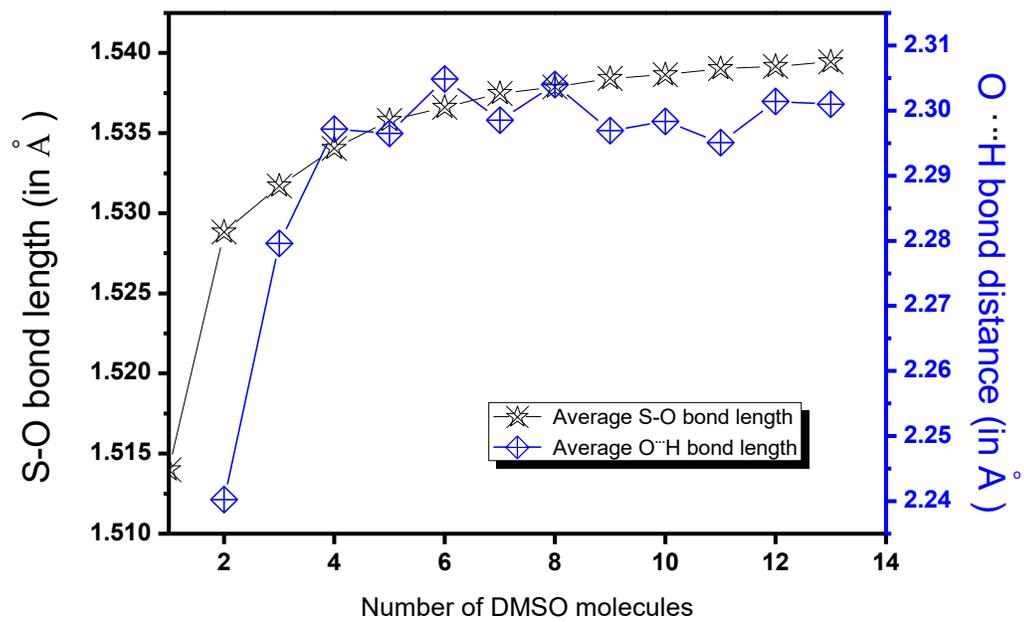

**Figure 3.**

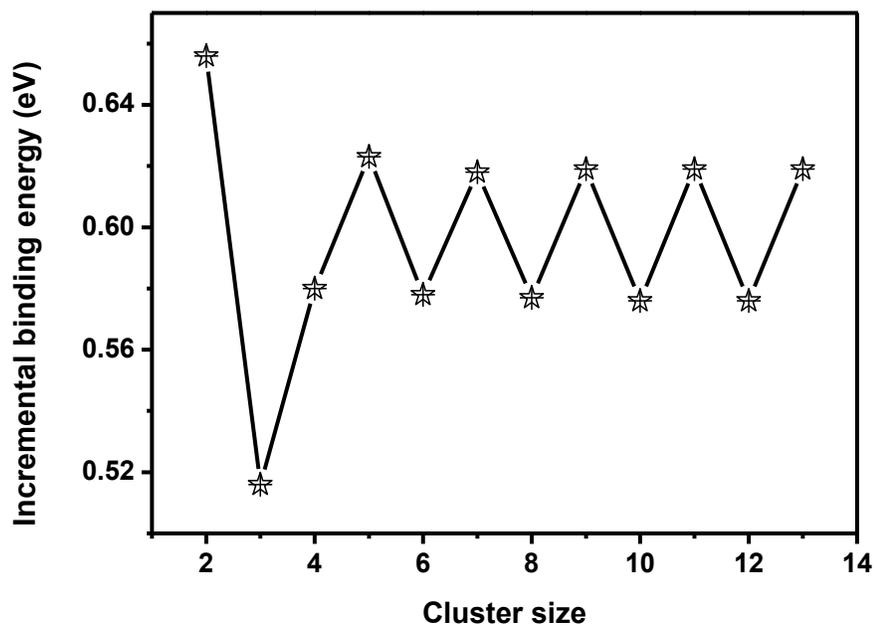

**Figure 4.**

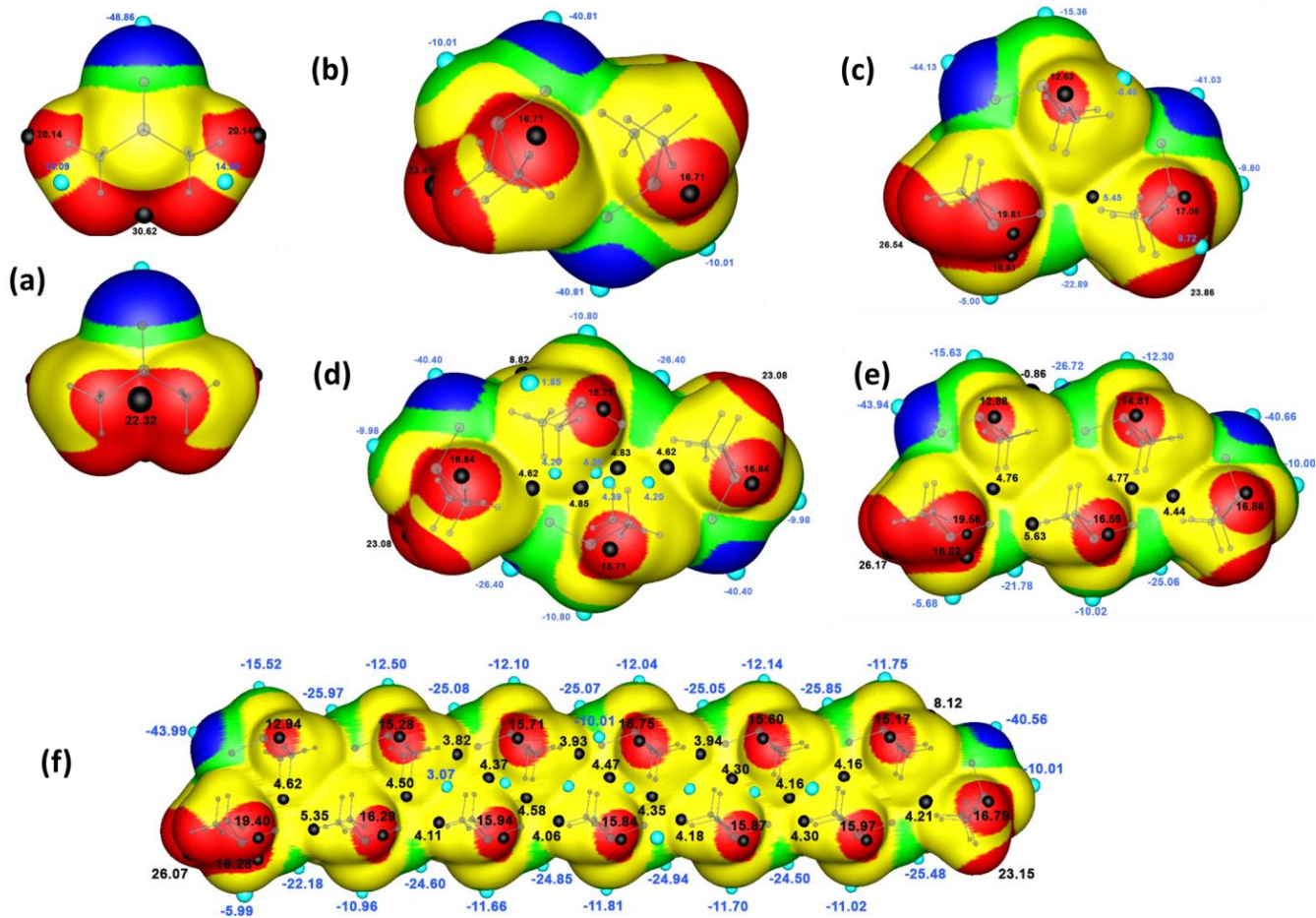

**Figure 5.**

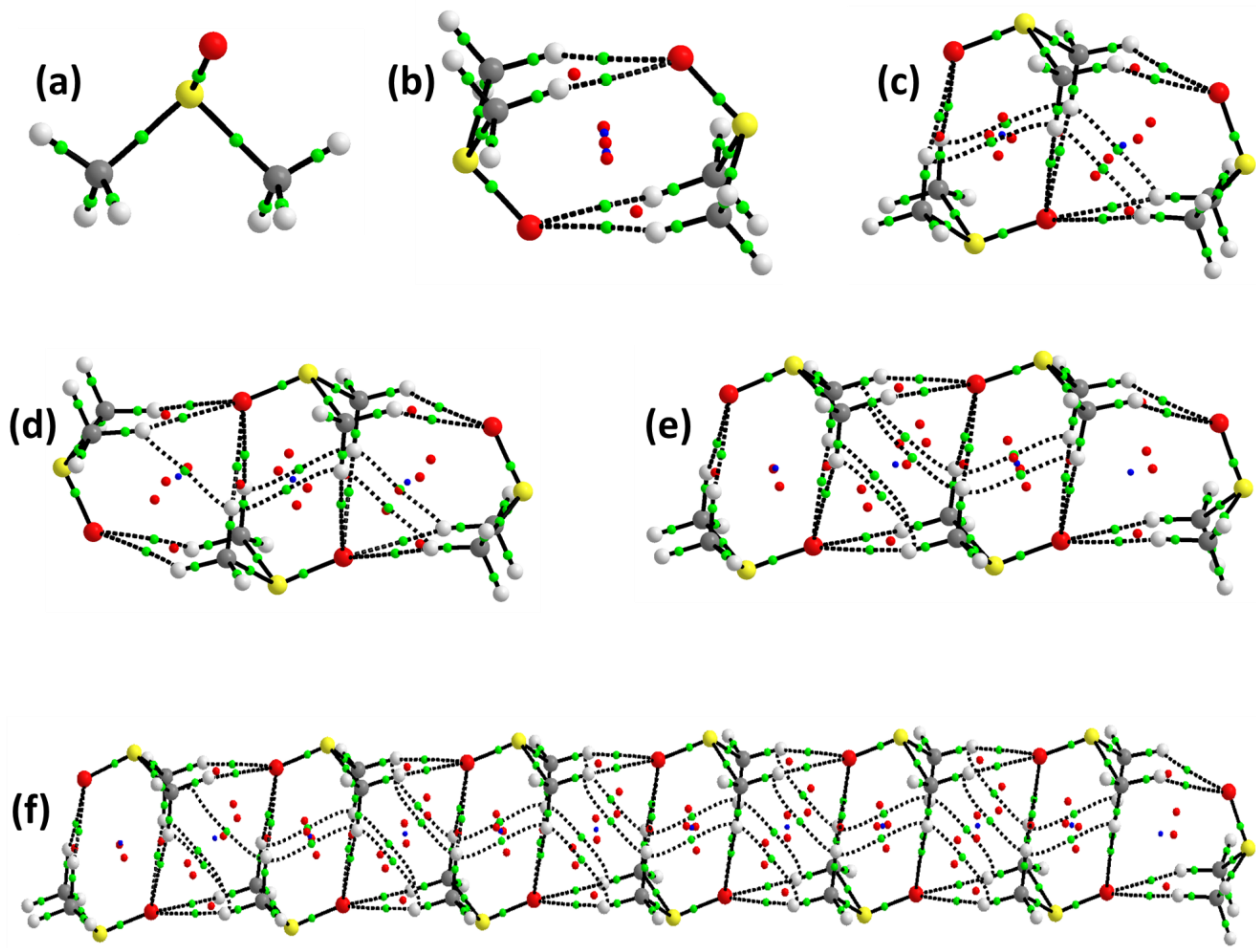

**Figure 6**

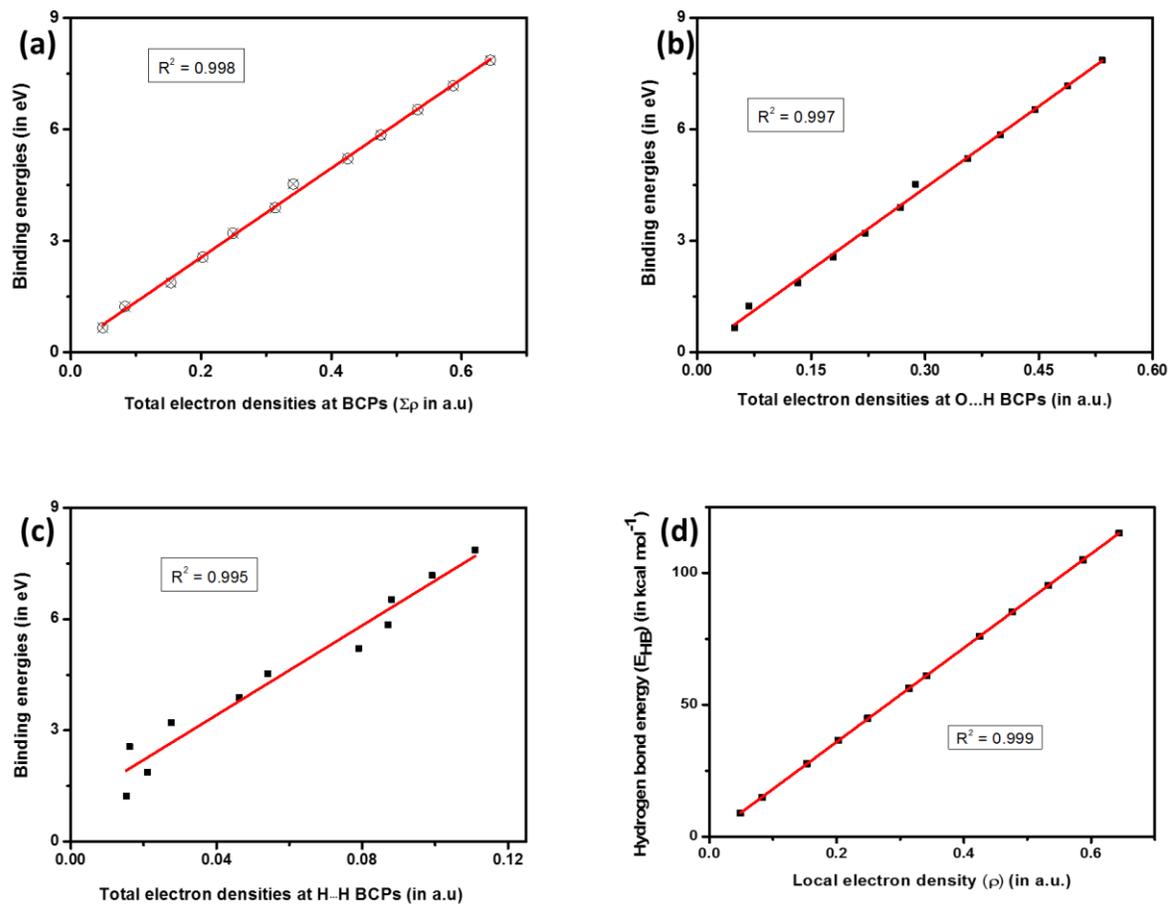

**Figure 7**

1   **Table 1.** Binding Energy (eV) of (DMSO)$_n$, n = 2 – 13, at various levels of theory using 6-311++G(d,p) basis set for the most stable

2   DMSO clusters optimized with the respective functional.

| Cluster size (n) | Binding Energy (eV) | | | | | | | | BE/n (eV) | IBE (eV) |
|---|---|---|---|---|---|---|---|---|---|---|
| | | | | B3LYP-D3 | | M062X-D3 | | | | |
| | B3LYP | M05-2X | M06-2X | Uncorrected | Corrected | Uncorrected | Corrected | MP2 | | |
| 2 | 0.410 | 0.648 | 0.677 | 0.656 | 0.706 | 0.694 | 0.757 | 0.614 | 0.328 | 0.656 |
| 3 | 0.715 | 1.187 | 1.237 | 1.232 | 1.348 | 1.286 | 1.428 | 1.180 | 0.411 | 0.516 |
| 4 | 1.089 | 1.800 | 1.874 | 1.871 | 2.051 | 1.956 | 2.178 | 1.818 | 0.468 | 0.580 |
| 5 | 1.489 | 2.445 | 2.541 | 2.560 | 2.811 | 2.659 | 2.965 | 2.516 | 0.512 | 0.623 |
| 6 | 1.850 | 3.050 | 3.171 | 3.201 | 3.520 | 3.324 | 3.713 | 3.581 | 0.534 | 0.578 |
| 7 | 2.256 | 3.701 | 3.846 | 3.885 | 4.275 | 4.034 | 4.510 | 3.819 | 0.555 | 0.618 |
| 8 | 2.617 | 4.301 | 4.471 | 4.524 | 4.982 | 4.696 | 5.256 | 4.453 | 0.566 | 0.577 |
| 9 | 3.022 | 4.958 | 5.151 | 5.209 | 5.738 | 5.409 | 6.055 | 5.148 | 0.579 | 0.619 |
| 10 | 4.533 | 5.560 | 5.779 | 5.848 | 6.445 | 6.072 | 6.804 | 5.785 | 0.585 | 0.576 |
| 11 | 3.788 | 6.215 | 6.458 | 6.532 | 7.200 | 6.787 | 7.604 | 6.478 | 0.594 | 0.619 |
| 12 | 4.149 | 6.818 | 7.087 | 7.171 | 7.906 | 7.450 | 8.352 | 7.114 | 0.598 | 0.576 |
| 13 | 4.554 | 7.473 | 7.765 | 7.856 | 8.662 | 8.164 | 9.152 | 7.809 | 0.604 | 0.619 |

**Table 2.** Pairwise counterpoise corrected interaction energies (kcal mol$^{-1}$) of the trimer, tetramer, pentamer, hexamer and Heptamer.

| Pair | Trimer | Tetramer | Pentamer | Hexamer | Heptamer |
|------|--------|----------|----------|---------|----------|
| 1,2 | -13.27 | -4.59 | -4.67 | -4.61 | -13.41 |
| 1,3 | -5.47 | -12.52 | -12.77 | -12.66 | 0.744 |
| 1,4 | – | -5.25 | 0.71 | 1.01 | 1.50 |
| 1,5 | – | – | -5.79 | 0.78 | 0.78 |
| 1,6 | – | – | – | -5.48 | 0.64 |
| 1,7 | – | – | – | – | -5.72 |
| 2,3 | -4.86 | -5.25 | -5.34 | -5.34 | -5.82 |
| 2,4 | – | 0.10 | 0.55 | 0.78 | 0.56 |
| 2,5 | – | – | 0.63 | 0.53 | 0.79 |
| 2,6 | – | – | – | 0.91 | 1.71 |
| 2,7 | – | – | – | – | 2.86 |
| 3,4 | – | -4.59 | -5.65 | -5.48 | -5.61 |
| 3,5 | – | – | 2.88 | 0.91 | 0.81 |
| 3,6 | – | – | – | 3.15 | 3.09 |
| 3,7 | – | – | – | – | -12.99 |
| 4,5 | – | – | -13.41 | -4.61 | -4.62 |
| 4,6 | – | – | – | -12.66 | -12.67 |
| 4,7 | – | – | – | – | 0.99 |
| 5,6 | – | – | – | -5.48 | -5.32 |
| 5,7 | – | – | – | – | 0.77 |
| 6,7 | – | – | – | – | -5.53 |

1 **Table 3**. The number of intermolecular bond critical points (BCPs), ring critical points (RCPs) and cage critical points (CCPs),  QTAIM
2 BCPs parameters corresponding to intermolecular bonding between the DMSO molecules.

| Cluster size (n) | BCPs/RCPs/CCPs | $\Sigma \rho$ (r) | $\Sigma \nabla^2 \rho$ (r) | $\Sigma \lambda_1$ | $\Sigma \lambda_2$ | $\Sigma \lambda_3$ | $\Sigma H$ (r) | - $\Sigma$ G(r)/$\Sigma$ V(r) |
|---|---|---|---|---|---|---|---|---|
| 2 | 4/5/2 | 0.04894 | 0.14979 | -0.05147 | -0.04685 | 0.24810 | 0.00452 | 4.637 |
| 3 | 10/12/2 | 0.08324 | 0.26888 | -0.07980 | -0.06765 | 0.41635 | 0.00987 | 12.624 |
| 4 | 17/17/3 | 0.15348 | 0.48903 | -0.14833 | -0.12696 | 0.76432 | 0.01707 | 20.861 |
| 5 | 22/22/4 | 0.20264 | 0.64779 | -0.19717 | -0.17190 | 1.01687 | 0.02282 | 27.249 |
| 6 | 27/28/5 | 0.24901 | 0.79734 | -0.24047 | -0.21148 | 1.24928 | 0.02808 | 33.326 |
| 7 | 36/36/6 | 0.31379 | 1.01024 | -0.30140 | -0.25982 | 1.57146 | 0.03666 | 45.151 |
| 8 | 40/42/7 | 0.34179 | 1.09698 | -0.32561 | -0.28033 | 1.70293 | 0.03979 | 50.163 |
| 9 | 50/50/8 | 0.42521 | 1.37355 | -0.40587 | -0.34792 | 2.12734 | 0.05052 | 63.027 |
| 10 | 56/56/9 | 0.47615 | 1.53866 | -0.45262 | -0.38992 | 2.38120 | 0.05654 | 70.467 |
| 11 | 63/63/10 | 0.53291 | 1.72314 | -0.50738 | -0.43530 | 2.66583 | 0.06358 | 79.467 |
| 12 | 70/70/11 | 0.58725 | 1.90103 | -0.55678 | -0.47770 | 2.93551 | 0.07036 | 88.361 |
| 13 | 77/77/12 | 0.64426 | 2.08660 | -0.61153 | -0.52243 | 3.22057 | 0.07750 | 97.397 |

3
4
5

2 **Graphical Abstract**

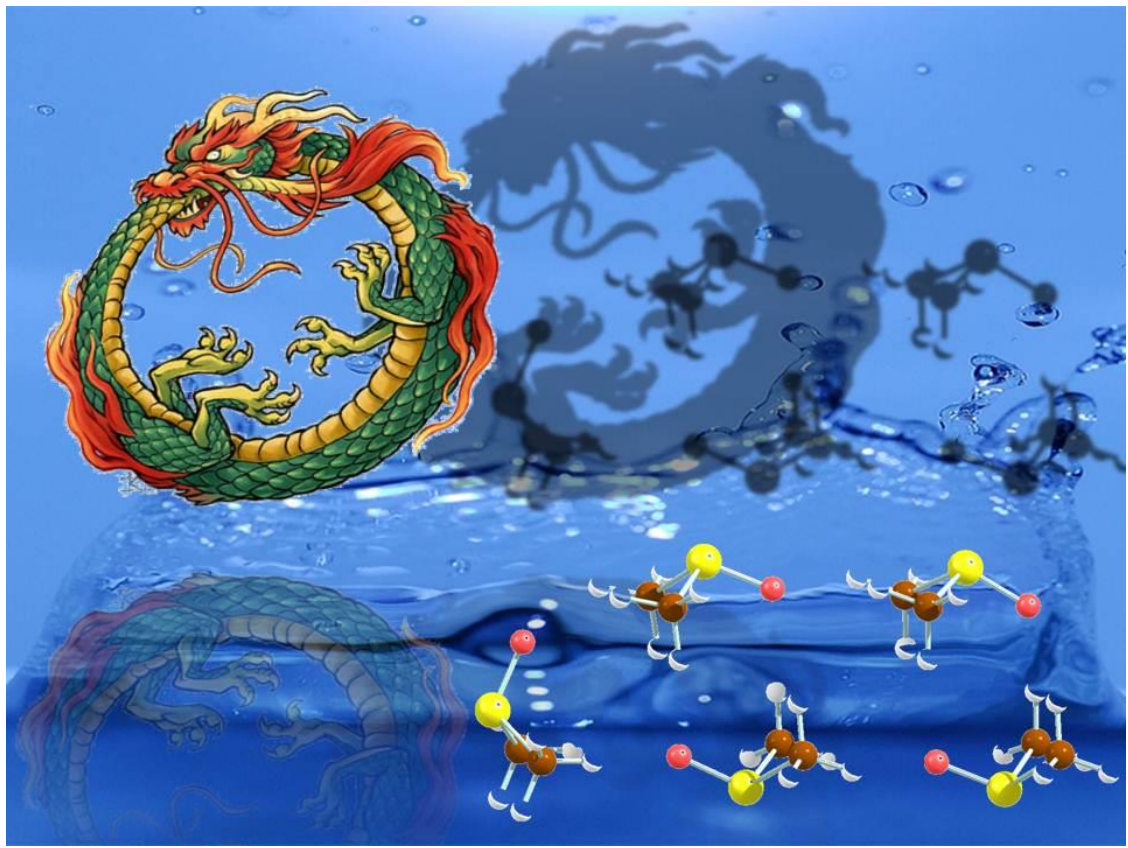

4
5